\documentclass[12pt,twocolumn]{aastex631}
\begin{document}

\title{ROME III. The Arecibo Search for Star-Planet Interactions at 5 GHz}

\author[0000-0001-6987-6527]{Matthew Route}
\affiliation{Department of Astronomy and Astrophysics, the Pennsylvania State University, 525 Davey Laboratory, University Park, PA 16802, USA}
\affiliation{Center for Exoplanets and Habitable Worlds, the Pennsylvania State University, 525 Davey Laboratory, University Park, PA 16802, USA}
\affiliation{Northrop Grumman Electronic Systems, 6120 Longbow Drive, Boulder, CO 80301, USA}
\affiliation{Department of Physics and Astronomy, the University of Mississippi, 121B Lewis Hall, University, MS 38677, USA}

\author[0000-0003-1915-5670]{Alexander Wolszczan}
\affiliation{Department of Astronomy and Astrophysics, the Pennsylvania State University, 525 Davey Laboratory, University Park, PA 16802, USA}
\affiliation{Center for Exoplanets and Habitable Worlds, the Pennsylvania State University, 525 Davey Laboratory, University Park, PA 16802, USA}

\correspondingauthor{Matthew Route}
\email{mproute@olemiss.edu}


\begin{abstract}
After nearly three decades of discovery, many exoplanetary systems have been studied and characterized in detail with one important exception: exoplanet magnetism.  Although many surveys sought to detect magnetospheric radio emissions from exoplanets to directly measure their magnetic field strengths, they have yet to reveal an unambiguous detection.  However, the indirect detection of exoplanet magnetic fields by measuring their influence on their host stars via magnetic star-planet interactions has recently gained prominence as an alternative method of discovery.  This third paper of the ROME (Radio Observations of Magnetized Exoplanets) series presents the results of a targeted radio survey of eight nearby exoplanet-hosting systems that may engage in star-planet interactions.  This survey, conducted with the Arecibo radio telescope at $\sim$5 GHz, has the greatest frequency coverage of any to date, while providing mJy-level sensitivity over $<$1 s integration times.  No exoplanet-induced stellar radio bursts were detected.  The orbital phase coverage of candidate systems for magnetic star-planet interactions is described, and the survey results are examined within the context of the plasma flow-obstacle paradigm and searches for star-planet interactions at other wavelengths.
\end{abstract}
 
\section{Introduction}
Stars interact with their orbiting exoplanet companions via a variety of mechanisms, but perhaps none is more controversial than star-planet interactions \citep{lan18}.  \citet{cun00} hypothesized that tidal and magnetic interactions between host stars and exoplanets with semimajor axes $a\lesssim 0.1$ au might increase stellar activity in several ways.  Tidal forcing of the stellar atmosphere and convective zone may increase local or global magnetic activity, while unipolar or dipolar magnetic plasma flow-obstacle interactions may enhance activity in stellar atmospheres \citep{zar18}.  Magnetic and tidal effects should be distinguishable by the presence of either one or two enhancements in activity per stellar rotation period, respectively.  Both tidal and magnetic activity enhancement should be most pronounced in the corona and transition region, and to a lesser extent the chromosphere, due to their comparatively lower densities and proximity to the interacting exoplanet companion \citep{saa01}.

Observational support for these interactions originally came from a chromospheric enhancement monitoring campaign of hot jupiter host stars including HD 179949, $\upsilon$ And, and HD 189733 \citep{shk05,shk08}.  Several data sets of Ca II K residuals appeared to show activity enhanced by $\sim$1\%, but since the modulation varied from epoch to epoch and typically did not occur at the orbital phases expected from initial theoretical models, a more complicated energy transport model was required to explain the intermittent activity.  Later calculations found that the detected chromospheric activity was several orders of magnitude greater than is available from plasma flow-obstacle interactions (e.g., see \citealt{zar07} on HD 179949) calling such detections into question.  Similarly, statistical analyses of stars with hot exoplanet companions came to conflicting conclusions, with some failing to find a statistically significant correlation between exoplanets with small semimajor axes and chromospheric activity enhancement (e.g., \citet{can11,pop14}).

\citet{rub00} hypothesized that as-yet-undetected planetary companions could cause the emission of superflares ($E\sim$10$^{33}$-10$^{38}$ erg) from F and G main-sequence stars.  They argued that the properties of the flares emitted from $\kappa$ Ceti and $\pi^{1}$ UMa were similar to those emitted by RS Canum Venaticorum (RS CVn) stars.  They proposed that open field lines from the host star were anchored at the exoplanetary companion, which would wrap the field lines about the host star as the exoplanet orbits it, thereby amplifying the stored magnetic energy.  Subsequent interactions of magnetic loops between the star and exoplanet would result in magnetic reconnection events that trigger superflares.  They further hypothesized that these flares would be luminous at X-ray to radio wavelengths, in analogy to RS CVn and solar flares.

More recently, attention has focused on the radio emission from M dwarfs that may result from star-planet interactions with potential exoplanet companions, including the M9 dwarf TVLM 513-46546 as observed at 4-5 GHz  \citep{wol14,cur20}, the M4.5 dwarf YZ Ceti at 550-900 MHz and 2-4 GHz \citep{pin23,tri23}, the M5.5 dwarf Proxima Centauri at 1.1-3.1 GHz \citep{per21}, the M4.5 dwarf GJ 1151 at 150 MHz \citep{ved20,bla23}, and several M1.5--M6.0 dwarfs observed at 144 MHz \citep{cal21}.  In each case, highly circularly polarized, high brightness temperature emission consistent with electron cyclotron maser (ECM) emission was detected that may result from magnetic star-planet interactions.  Alternatively, the observed characteristics at lower frequencies are also consistent with plasma radiation, and several other non-planet-hosting M dwarfs exhibit similar emission (e.g., \citet{ste01,lyu20})

Therefore, radio surveys are a potentially powerful tool to observe coronal flaring and shock processes that may result from magnetic star-planet interactions and diagnose their associated plasma density, local magnetic field, particle acceleration, and wave propagation processes (e.g., \citet{ben10}).  While magnetic star-planet interactions may cause detectable stellar radio emissions, it is important to note that they would do so against a backdrop of a legion of intrinsic stellar radio phenomena.  High spatial and temporal resolution observations of the Sun reveal that thermal bremsstrahlung, incoherent gyrosynchrotron, and coherent emissions processes contribute to its myriad radio emissions.  At mm to cm wavelengths, gyrosynchrotron dominates, while at dm to m wavelengths, coherent plasma radiation dominates \citep{bas98}.  The ECM emission may cause the fine structure within type IV bursts and solar microwave spikes, which range in frequency from $\nu\sim$50 MHz to 10 GHz and $\nu\sim$100 MHz to 5 GHz, respectively \citep{tre06,pic08}.  \citet{dul87} summarized how these four radio emission mechanisms - thermal bremsstrahlung, gyrosynchrotron, plasma, and ECM - differ in brightness temperature, circular polarization fraction, frequency of operation, and timescale of variability.  A search for stellar flaring at 5-10 GHz may be especially promising, since gyrosynchrotron microwave bursts tend to peak within this frequency window, and several types of ECM-generated activity are also present.

Given these considerations, we conducted a survey of nearby exoplanet-hosting systems at Arecibo Observatory (AO) in 2010-2011 to search for enhanced coronal radio emissions that may be linked to magnetic star-planet interactions.  A detailed account of the analysis and interpretation of one exoplanet system that we surveyed, HD 189733, was presented in ROME I and II \citep{rou19,rl19}.  Section 2 describes our target selection methodology, instrumentation, and observations.  Section 3 presents the detection limits and orbital phase coverage of our survey for potential magnetic star-planet interactions.  Section 4 contextualizes these results with respect to searches for magnetic star-planet interactions at other wavelengths and considers the reasons for nondetection. Section 5 concludes by reviewing the significance of this survey and providing suggestions for future work.  The results of our survey indicate that the detection and characterization of exoplanet magnetism at radio wavelengths via magnetic star-planet interactions remains an admirable, although difficult, goal.

\section{Target Selection and Observations}

\subsection{Target Selection}

Target systems were chosen based on the potential for their physical properties to generate magnetic star-planet interactions and their location in the sky (Table 1).  Physical properties that served as selection criteria include large masses and small semimajor axes, $a$.  Exoplanet mass may alter the potential for star-planet interactions in somewhat different ways depending upon whether a dipolar or unipolar magnetic interaction with the host star occurs.  For a strongly magnetized exoplanet, dipolar interactions would cause nearly continuous reconnection between large- and/or small-scale stellar  magnetic fields and the exoplanetary magnetic field.  In this case, theoretical models indicate that an exoplanet with a larger mass would have a larger convective core and greater thermal flux to drive a magnetic dynamo that results in a stronger magnetic field \citep{gri07,chr09,rei10}.  A stronger magnetic field would yield a larger magnetosphere (obstacle) to interact with a strongly magnetized stellar wind (plasma flow).  On the other hand, unipolar interactions between a weakly/ unmagnetized exoplanet and a strongly magnetized stellar wind would create a field-aligned current circuit propagated via magnetohydrodynamic waves.  In this case, a more massive exoplanet presents a larger cross-section (obstacle) via its surface or its ionosphere to interact with a strongly magnetized stellar wind (plasma flow; \citealt{zar18}).  Exoplanets close to their host stars may be tidally locked with weakened or nonexistent intrinsic magnetic fields \citep{zar07}.  In both cases, $\gtrsim$keV electrons would be accelerated toward the source of the strong magnetic field, the magnetized star, with a fraction of the electronic kinetic energy radiated away at radio wavelengths.  Each targeted system has at least one exoplanet at a distance $a\lesssim0.1$ au from its host star, which is the range considered by \citet{cun02} to be especially promising to enhance stellar activity. Our target stars range from spectral types G0 to M2.5 and include 51 Peg and HD 209458, which both appeared in their study, as well as GJ 176, HD 46375, 55 Cnc, GJ 436, HD 102195, and HD 189733.

Physical system properties aside, we targeted systems $<$50 pc from the Sun.  Since stellar coronal flare luminosities follow a power-law distribution (e.g., \citealt{gud03}), increasingly distant sources require stronger and rarer magnetic star-planet interaction induced flares for detection.  The single, fixed dish nature of the Arecibo radio telescope also constrained our targets to have declinations of 0$^\circ$ to +38$^\circ$.

\subsection{Instrumentation and Observations}

The targeted exoplanet systems were surveyed during AO program A2471, conducted from 2010 January 6 to 2011 September 7.  Each exoplanetary system was observed for a total of 1-2 hr, sometimes across multiple epochs, although no single epoch was longer than 2 hr (Table 2\footnote{Data sets referenced in this table may be retrieved upon request from the Arecibo Observatory Tape Library, to be hosted by Texas Advanced Computing Center (TACC, https://www.tacc.utexas.edu) after 2023 Aug 14.}).  The $\sim$2 hr observing session limitation represents the maximum time for a target to transit the field of view of the fixed dish of the Arecibo radio telescope.  During these observing sessions, 20 s calibration on-off scans using a local oscillator were interlaced between 10 minute target scans.  

Although the 305 m dish is fixed, the William E. Gordon radio telescope at AO offered several advantages relative to other facilities, including its exquisite sensitivity, accurate polarization measurements over short time scales, and large, $\sim$1 GHz bandpass, which was exemplary in 2010-2011.  The observing system consisted of the C-band receiver\footnote{``C-Band,'' available at http://www.naic.edu/$\sim$astro/RXstatus/Cband/Cband.shtml} and the recently commissioned Mock spectrometer\footnote{``The Mock Spectrometer,'' http://www.naic.edu/$\sim$astro/mock.shtml}.  The system temperature ranged from 25--32 K, with antenna gains of 5.5--9.0 K Jy$^{-1}$.  The half-power beamwidth varied from 0.97' $\times$ 1.09' (azimuth $\times$ zenith angle) at 4.500 GHz central frequency to 0.79' $\times$ 0.92' at 5.400 GHz (Chris Salter, personal communication).  Full Stokes parameters were computed by the Mock spectrometer array based on the dual-linear polarization input signals from the receiver.  This array consists of seven field-programmable gate array (FPGA) Fast Fourier Transform (FFT) Mock spectrometers, which were individually tuned to central frequencies of 4.325, 4.466, 4.608, 4.750, 4.892, 5.034, and 5.176 GHz, resulting in an $\sim$1 GHz simultaneous bandpass stretching from 4.239 to 5.262 GHz.  Each 8192-channel spectrometer yields a 172 MHz bandpass that overlaps the frequency range of each neighboring spectrometer by $\sim$30 MHz.  The spectrometers were configured for 0.1 s sampling.

Prior to analysis, data were rebinned to a frequency (time) resolution of 83.6 kHz (0.9 s) to improve the signal to noise ratio characteristics.  Radio frequency interference (RFI) was reduced through an iterative statistical process detailed in \citet{rouphd}.  Data analysis consisted of the creation of dynamic spectra (time-frequency spectrograms) and bandpass-averaged time series graphs of flux density for each Stokes parameter in each spectrometer.  Since one of the hallmarks of the two most likely types of radio emission, ECM and gyrosynchrotron, is significant circular polarization, we searched Stokes V dynamic spectra for $\gtrsim$10\% bursts of emission, which are readily apparent when present (e.g., Figure 1).  AO is insensitive to quiescent radio emission due to its local calibration procedure and confusion limitations; thus, only rapid ($\sim$minutes), circularly polarized radio bursts could be detected. Candidate bursts were identified by surpassing a 3$\sigma$ noise threshold in their Stokes V time series graph.  The 1$\sigma$ instrumental sensitivity was $\sim$0.1-1.2 mJy, with per-scan sensitivity distribution properties described in \citet{rou17}.  Once identified, candidate burst Stokes Q and U components were examined for behavior indicative of RFI, and their morphology was assessed against a library of RFI artifacts acquired during our previous surveys at AO to eliminate false positives (Figure 1).

\section{Results}
No radio bursts of any type were detected from any exoplanet-hosting system in our survey.  We emphasize that this survey was sensitive only to $\gtrsim$10\% circularly polarized bursts of ECM or gyrosynchrotron radio emission of several minutes duration and was entirely insensitive to quiescent (unpolarized and/or slowly varying) emission.  However, we can compute 3$\sigma$ detection limits on the radio emission from target stars using the standard deviation ($\sigma$) of the frequency-integrated time series of the cleanest Mock spectrometer, centered at 4.466 GHz (Table 3).  Comprehensive, frequency-dependent sensitivity limits on a system-by-system basis are available in \citet{rouphd}.  A comparison of these detection limits to several types of stellar flares is depicted in Figure 2.

Incidentally, since each surveyed planetary system fits entirely within the half-power beamwidth, not only do these detection limits constrain the luminosity of stellar flares induced by magnetic star-planet interactions, but they also constrain the detectability of magnetospheric radio emissions from exoplanets within those systems.  Planetary systems with massive exoplanets close to their parent stars are thought to be especially promising targets for the detection of magnetospheric, or auroral, planetary radio emission due to the larger incident Poynting flux and denser stellar wind impinging on their magnetospheres \citep{zar18}.  Although this survey failed to detect any such radio emissions, the detection limits in terms of maximum Jovian ECM-induced emissions are also reported in Table 3.

\subsection{Orbital Phase Coverage of Companions for Star-Planet Interactions}

One important consideration in our search for magnetic star-planet interactions is the orbital phase coverage of our observations.   Borrowing from the nomenclature for hot Jupiters (e.g., \citet{daw18}), we compute the orbital phase coverage for the hot ($P<$10 days) exoplanet in each system,

\begin{equation} T_{0} = T_{t} + P~E_{Orb},\end{equation}

where $T_{0}$ represents the beginning or end of the observation, $T_{t}$ represents the measured (or estimated) midpoint of transit (or periapse), $P$ is the exoplanet orbital period, and $E_{orb}$ is the orbital cycle.  Orbital phase 0.0 denotes when transit or periapse would occur.  The results for hot exoplanets are shown in Table 4 and Figure 3.  

Even among the shortest-period exoplanets, the orbital phase coverage is quite modest since many targets were observed for $\sim$1 hr, and only four targeted systems were observed for $>$1.5 hr.  The peak orbital phase coverage occurs for 55 Cnc e ($\Delta\phi_{orb}$=0.08), although the coverage and timing of the observations for HD 189733 are also significant (Section 4.1.2).  We can also calculate the orbital phase coverage for the warm (10 days$\leq~P<$200 days) exoplanet 55 Cnc b, which is a tiny $\Delta\phi_{orb}$=0.004.  Thus, despite the excellent sensitivity of AO, modest orbital phase coverage only loosely constrains the incidence of star-planet interactions within the surveyed systems.
 
\section{Discussion}

\subsection{Exoplanet Systems of Special Interest}

\subsubsection{Targets with Previous Magnetic Star-Planet Interaction Observations}

\emph{HD 46375.--} \citet{shk05} searched for periodic star-planet induced chromospheric variability in the \ion{Ca}{2} H and K lines (3968 and 3933 $\AA$) among HD 46375 and several other program stars.  No change was detected to a level of 0.001.  Similarly, we did not detect any sign of magnetic star-planet interactions in this system, although our observations only occur at $\phi_{orb}\sim$0.68. This orbital phase is somewhat removed from $\phi_{orb}\sim$0.5, which theory indicates may experience augmented activity due to tidal star-planet interactions.

\emph{55 Cnc.--} The 55 Cnc binary system consists of a K0 primary, an M4.5 secondary, and five exoplanets that circle the primary.  The closest orbiting exoplanet, the sub-Neptune 55 Cnc e, may be close enough to its host star to cause star-planet interactions, although its magnetosphere would likely present a very small cross section to the stellar wind for plasma flow-obstacle interactions.  While \citet{fol20} determined that 55 Cnc e likely orbits within its parent star's Alfv\'{e}n radius, \citet{mor21} determined that a planetary magnetic field of $B\sim$1 G would generate $\sim$10$^{5}$ less energy than required to explain the photometric phase variations observed in the system by CHEOPS.  Similarly, we found no evidence for star-planet induced coronal emissions in this system.

\emph{51 Peg.--} \citet{saa01} found no sign of star-planet interactions in the \ion{Ca}{2} infrared triplet (IRT) down to $\sim$3.2\%.  \citet{shk03} reported significant nightly activity for 51 Peg which they suggested might be cyclical and phase-shifted based on $\sim$10 nights of \ion{Ca}{2} H and K observations.  Subsequent observations found no variability down to 0.001 \citep{shk05}.  We found no evidence for magnetic star-planet induced coronal activity either at $\phi_{orb}\sim$0.22 or at $\phi_{orb}\sim$0.43, near the anticipated tidal activity enhancement zone.

\emph{HD 209458.--} As with 51 Peg b, preliminary monitoring of chromospheric \ion{Ca}{2} H and K activity suggested nightly activity which could be periodic in nature, but was later found to lack variability down to the level of 0.001 \citep{shk03,shk05}.  Alternatively, \citet{jen12} found an anomalous shifting of the broad component relative to the narrow component of the H$\alpha$ line in transmission spectra during transit and secondary eclipse.  They suggested that stellar chromospheric activity enhanced by star-planet interactions may alter the H$\alpha$ line shape, although this phenomenon may also point to a data reduction artifact.  Our radio observations do not find evidence of magnetic star-planet induced coronal activity within this system at $\phi_{orb}\sim$0.01 or $\phi_{orb}\sim$0.87.  Note that enhanced activity near $\phi_{orb}=$0.0 would be anticipated in both tidal and magnetic interaction cases.

\subsubsection{HD 189733}	

The proximity of the HD 189733 system to the Earth, the brightness of the primary star, and the fortuitous alignment of its hot Jupiter companion that results in a transit every 2.219 days have made the system a particularly inviting target for study.  With a chromospheric activity index, S-value=0.525, HD 189733 A is among the 10\% most active K dwarfs in the the California and Carnegie Planet Search Project.  Transmission spectroscopy of the frequent transits has resulted in the discovery of the exoplanet atmosphere's composition, thermal properties, and evaporation.  Multiwavelength observations of the system have revealed a wealth of stellar activity from optical to X-ray wavelengths, including spots and flares at all rotational and exoplanetary orbital phases.  Compellingly, several research groups announced the discovery of enhanced chromospheric, transition region, and coronal activity that may be phased with exoplanet orbital phases $\phi_{orb}\sim$0.5--1.0 and caused by magnetic and/or tidal star-planet interactions.  In one case, it was furthermore suggested that the primary star may actively accrete evaporated material from HD 189733 b (see ROME I and references therein).

In ROME I, we analyzed AO radio observations to search for enhanced coronal activity associated with such star-planet interactions.  We also analyzed \emph{Microvariability and Oscillations of Stars (MOST)}, Automated Photoelectric Telescope (APT), and Wise photometry to search for periodicities suggestive of stellar hemispheres that contain more darkened starspots than average, which may point to enhanced photospheric activity associated with these interactions.  We also compiled all observations and reported instances of stellar spotting and flaring at radio through X-ray wavelengths to attempt to discover enhancements phased to the orbital period of the exoplanet.  Our results could find no statistically significant enhancement, including at the orbital phases where other research groups thought star-planet interactions may occur.  In ROME II, we again leveraged \emph{MOST}, APT, and Wise photometry to search for photometric bright features that may be related to accretion structures, using the T Tauri paradigm as a template.  After computing plasma parameters such as density, temperature, accretion rate, and flare structure sizes, we determined an accretion rate of $\dot{M}\sim10^{9}$ to 10$^{11}$ g s$^{-1}$, which would result in undetectably low activity at all wavelengths.  Since the publication of ROME II, new flare observations and several reanalyses of existing optical and UV data sets have been presented, which we will address in a forthcoming publication (e.g., \citet{bou20} and references therein).

However, for the purposes of this paper, we focus on research that may directly influence the ability of HD 189733 A or its exoplanet companion to produce radio emission.  Based on their convective dynamo model, \citet{rei10} estimated that HD 189733 b would have a detectable radio flux density of $\Phi_{radio}\sim$57 mJy up to an ECM cutoff frequency of $\nu\sim$39 MHz.   \citet{zag18} used a revised interior conductivity profile to estimate peak ECM emission at $\nu\sim$15 MHz with a radio flux density of $\Phi_{radio}\sim$20 mJy.  \citet{kav19} leveraged the 3D magnetohydrodynamic modeling code BATS-R-US to investigate the prospects for the detection of radio emissions from HD 189733 b. They used the magnetic field topology maps of HD 189733 A derived from  several epochs of Zeeman Doppler Imaging (ZDI) observations from \citet{far17} as inputs.  They found that an exoplanetary polar magnetic field of $B\sim$10 G would be required to create ECM emission at $\nu\sim 25$ MHz that is higher than the plasma frequency of the stellar wind environment and thus could escape to an observer with $\Phi_{radio}\sim$50-130 mJy.  Our radio observations do not shed any light on these predictions, as no exoplanet magnetospheric activity was observed.  We also find no evidence for magnetic star-planet induced coronal activity within this system at $\phi_{orb}\sim$0.57--0.61 (ROME I).

\subsection{Potential Reasons for Nondetection of Star-Planet Interactions}
There are several potential reasons for the nondetection of coronal radio emissions induced by magnetic star-planet interactions, including sensitivity, observing frequency, observation duration, and viewing geometry.  Considering sensitivity first, ECM-induced stellar flaring would likely be detectable in all cases (Figure 2), as was found in our study comparing the activity of HD 189733 A to active M dwarf ECM flares (ROME I).  Similar to the findings of that work, only luminous gyrosynchrotron flares might be visible from some targets within our survey, as calculated from the G\"{u}del-Benz relationship and their measured X-ray emissions.  It is also possible that the radio counterparts of extremely luminous optical superflares may be visible from our survey \citep{sch00}.  As mentioned in Section 1, it has been hypothesized that such flares may be powered by the magnetic interaction of close-in giant exoplanets.  However, it is nontrivial to estimate the radio luminosity of flares based on their optical emissions alone (e.g., \citealt{pri14}).  If such flares have radio luminosities comparable to their optical emission, $L\sim$5$\times$10$^{31}$ erg s$^{-1}$, they would be readily detectable by our survey.

Figure 2 indicates that a single order-of-magnitude increase in sensitivity should make the stellar flares of the nearest active stars, such as HD 189733 A, accessible.  An improvement in sensitivity of $\sim$10$^{1}$--10$^{2}$ would also enable the characterization of coronal activity from older, less active stars such as the Sun.  However, given the difficulty of untangling star-planet induced flaring from intrinsic coronal activity (i.e. ROME I), searches for the former phenomenon will require high-cadence, multiepoch observations.

Next we consider the frequency range of our survey with respect to coronal magnetic fields.  While the mean solar photospheric magnetic field is $B\sim$3-10 G, within active regions, this field can reach $B\sim$100 G, and within sunspot umbrae, field strengths as high as $B\sim$6 kG have been measured \citep{pri14}.  Within the solar corona, magnetic fields as high as $B\sim$2.5 kG have been measured above photospheric active regions.  Gyroresonance, gyrosynchrotron, and ECM emissions all contribute to solar coronal radio emissions that range in frequency from $\nu=$1 to 20 GHz \citep{asc05}.  Among these, only activity caused by the latter two processes would be detectable at great distances.  Thus, the frequency range of our survey was well suited for the detection of stellar coronal radio emissions associated with both intrinsic and star-planet induced activity.

Companion exoplanets may induce enhanced photospheric, chromospheric, and coronal emissions, including radio emissions, in their host stars either once per orbit due to magnetic effects or twice per orbit due to tidal effects.  Thus, the host star rotation, hot exoplanet orbital, and system beat periods are all of interest to untangle the effects of intrinsic stellar activity from activity induced by star-planet interactions.  However, hot Jupiter orbital periods are shorter than their host star rotation periods, making the orbital phase coverage the timescale of interest.  As described in Section 3.1, the maximum orbital phase coverage for our hot Jupiter targets is $\Delta\phi_{orb}\leq$0.08 (Table 4).  Thus, to untangle the effects of magnetic star-planet induced activity from intrinsic stellar activity would require nearly complete orbital phase coverage over multiple orbits.  As ROME I and ROME II indicated, even for as well-studied a system as HD 189733, it remains unclear whether star-planet interactions occur there.

Finally, we consider the system viewing geometry.  Gyrosynchrotron radiation, ubiquitous in solar flares, is generated by mildly relativistic electrons with Lorentz factors of $1\lesssim\gamma\lesssim 5$.  Emission is increasingly beamed at higher energies, thereby reducing the opening angles of emission from 90$^\circ$ to 10$^\circ$ with respect to the local magnetic field, which can be quite complicated in active regions \citep{bas98}. In situ measurements from the magnetized solar system planets and maser numerical simulations indicate that ECM emission is beamed into a hollow cone with half-apex angles of 30$^\circ$ to 90$^\circ$ \citep{zar98,tre06}, while numerical simulations of ultracool dwarf ECM emission suggest that half-apex angles of $\sim$1$^\circ$ may be possible \citep{lyn15}.  This beaming pattern could critically constrain the detection of such emission from active regions participating in star-planet interactions.  However, given that stellar active regions likely span the entire stellar surface or may be restricted to large activity bands (e.g., \citet{mci14}) visible from many vantage points, it is unlikely that viewing geometry plays a major role in reducing star-planet interaction detectability.  On the other hand, the beaming of ECM emission may contribute to the nondetection of magnetospheric radio emissions from the exoplanet companions in the surveyed systems.

\section{Conclusion}

In this third installment in the Radio Observations of Magnetized Exoplanets (ROME) series, we step back from analyzing the details of a particularly interesting exoplanetary system that may engage in star-planet interactions, HD 189733 A/B/b (ROME I and II), and present the complete results of our 2010-2011 survey that sought to detect and characterize stellar radio emissions that may be caused by magnetic star-planet interactions with close-in exoplanets.  These interactions may induce enhanced coronal, transition region, chromospheric, and photospheric activity in their host stars.  Such activity may generate powerful ECM or gyrosynchrotron flares phased to the exoplanet orbit. 

The survey targeted eight systems that have exoplanet companions within $a\lesssim$0.1 au of the host star.  This survey leveraged the exquisitely sensitive 305 m Arecibo radio telescope tuned to 4.2-5.2 GHz to search for strong stellar magnetic fields and probe the optimal frequency range for the detection of ECM and gyrosynchrotron solar radio flares.  Dynamic spectra in all four Stokes parameters from the Mock spectrometers were used to search for $\gtrsim10$\% circularly polarized ECM and gyrosynchrotron emission and to mitigate RFI.  All targets were observed for $\lesssim$2 hrs, the maximum time for them to transit the fixed dish.

Our survey did not detect stellar radio flares induced by magnetic star-planet interactions.  The 3$\sigma$ sensitivity for targeted systems ranged from 0.98 to 2.1 mJy, resulting in detection limits of $\nu~L_{\nu}=5.1\times$10$^{23}$ erg s$^{-1}$ to 1.4$\times$10$^{25}$ erg s$^{-1}$.  These luminosity detection thresholds correspond to modest ECM flares, such as those that appear on the M3.5 AD Leo, and powerful gyrosynchrotron flares, such as from the young active star $\epsilon$ Eri.  It has been hypothesized that extremely luminous optical superflares from FGK stars might be caused by star-planet interactions, which have luminosities $L\sim$5$\times$10$^{31}$ erg s$^{-1}$.  Although our survey was sensitive to radio emissions of energy output equal to these optical superflares, it is unclear whether the \emph{radio} counterparts to such flares would be detectable with current instrumentation.  Incidentally, these luminosity detection thresholds also constrain the high-frequency magnetospheric (auroral) emissions that may come from the exoplanets themselves.

Several targets, including HD 46375, 55 Cnc, 51 Peg, HD 209458, and HD 189733, have previously been investigated for signs of chromospheric and photospheric activity enhancement due to potential tidal or magnetic star-planet interactions.  Yet in each case, subsequent work challenged the initial putative discoveries of increased magnetic activity phased with the orbital or beat periods of their hot Jupiter companions.  Similarly, our search for enhanced coronal activity did not detect signs of magnetic star-planet interactions.  The inability to detect intrinsic stellar flaring, much less enhanced coronal activity from star-planet interactions at radio wavelengths, indicates that future efforts will require an $\sim$10$^{1}$--10$^{2}$ increase in sensitivity.  

These survey results provide guidance as to how to enhance the scientific return of future surveys for star-planet interactions at radio wavelengths.  First, increases in sensitivity of 1-2 orders of magnitude are required to detect magnetic star-planet interactions.  Second, future surveys should simultaneously observe as wide a bandpass as possible given the great uncertainty in flare magnetic field strengths that may be generated by star-planet interactions.  Third, future surveys should include polarimetry to distinguish among and characterize various types of astrophysical emission, as well as an RFI mitigation strategy.  Fourth, multiepoch, high-cadence observations are required to untangle the effects of intrinsic stellar activity on differentially rotating stars from enhanced activity induced by tidal and magnetic star-planet interactions.  Finally, high-fidelity theoretical simulations are required to model the influence of star-planet interactions on stellar atmospheres and assess the prospects for observing enhanced photospheric, chromospheric, transition region, and coronal activity within hot exoplanet systems.
 
\section{Acknowledgments}

M.R. would like to acknowledge support from the Center for Exoplanets and Habitable Worlds and the Zaccheus Daniel Fellowship. The Center for Exoplanets and Habitable Worlds is supported by Pennsylvania State University and the Eberly College of Science.  Data storage and analysis support has been made possible with the Theodore Dunham, Jr. Grants for Research in Astronomy.  At the time of the observations that are the subject of this publication, the Arecibo Observatory was operated by SRI International under a cooperative agreement with the National Science Foundation (AST-1100968), and in alliance with Ana G. M\'{e}ndez-Universidad Metropolitana, and the Universities Space Research Association.

This research has made use of NASA's Astrophysics Data System and the SIMBAD database, operated at CDS, Strasbourg, France.  Guidance on properties and publications related to objects of interest were obtained from the catalog located at exoplanet.eu.

\facility{Arecibo}
\software{IDL, MATLAB}

\clearpage

\begin{deluxetable}{lllllllll}
\tabletypesize{\scriptsize}
\tablecolumns{9}
\tablewidth{0pt}
\tablecaption{Survey Target Properties}
\tablehead{
	\colhead{Name}&
	\colhead{R.A.}&
	\colhead{Dec.}&
	\colhead{Host Star} &
	\colhead{Dist.}&
	\colhead{Semimajor}&
	\colhead{Period}&
	\colhead{Mass}&
	\colhead{Property}\\
	\colhead{}&
	\colhead{({hh}\phn{mm}\phn{ss})}&
	\colhead{(\phn{\arcdeg}~\phn{\arcmin}~\phn{\arcsec})}&
	\colhead{Type}&
	\colhead{(pc)}&
	\colhead{Axis (AU)}&
	\colhead{(d)}&
	\colhead{(M$_{J}$)}&
	\colhead{Refs}
}
\startdata
GJ 176 b\tablenotemark{a} & 04 42 56 & +18 57 29 & M2.5 Ve & 9.49 & 0.066 & 8.7836 & 0.0264 & \underline{{\bf {\em 1}}},1\\
HD 46375 b & 06 33 13 & +05 27 47 & K1 IV & 29.52 & 0.0398 & 3.024 & 0.226 & {\bf 2},\underline{{\em 3}},3\\
55 Cnc e & 08 52 36 & +28 19 51 & K0 V & 12.59 & 0.0154 & 0.7365 & 0.027 & {\bf 4},\underline{5},{\em 6},6\\
$\phantom{55 Cncx}$b & $\phantom{08 52 36}$ & $\phantom{+28 19 51}$ & $\phantom{K0 V}$ & $\phantom{12.59}$ & 0.1134 & 14.6516 & 0.8036 & {\bf 4},\underline{\em 5},5\\
$\phantom{55 Cncx}$c& $\phantom{08 52 36}$ & $\phantom{+28 19 51}$ & $\phantom{K0 V}$ & $\phantom{12.59}$ & 0.2373 & 44.3989 & 0.1611 & {\bf 4},\underline{\em 5},5\\
$\phantom{55 Cncx}$f & $\phantom{08 52 36}$ & $\phantom{+28 19 51}$ & $\phantom{K0 V}$ & $\phantom{12.59}$ & 0.7708 & 259.88 & 0.1503 & {\bf 4},\underline{\em 5},5\\
$\phantom{55 Cncx}$d & $\phantom{08 52 36}$ & $\phantom{+28 19 51}$ & $\phantom{K0 V}$ & $\phantom{12.59}$ & 5.450 & 5574.2 & 3.12 & {\bf 4},\underline{\em 5},5\\
GJ 436 b & 11 42 11 & +26 42 24 & M2.5 V & 9.78 & 0.02887 & 2.644 & 0.0737 & {\bf 7},\underline{\em 8},8\\
HD 102195 b & 11 45 42 & +02 49 17 & G8 V & 29.36 & 0.049 & 4.1139 & 0.46 & \underline{\bf 9},{\em 10},10\\
HD 189733 b & 20 00 44 & +22 42 39 & K2 V & 19.78 & 0.0313 & 2.219 & 1.13 & \underline{{\bf {\em 11}}},12\\
HD 209458 b & 22 03 11 & +18 53 04 & G0 V & 48.15 & 0.0467 & 3.5247 & 0.714 & {\bf 13},\underline{{\bf 14}},{\em 8},8\\
51 Peg b & 22 57 28 & +20 46 08 & G2 IV & 15.53 & 0.053 & 4.231 & 0.46 & {\bf 15},\underline{{\em 3}},16\\
\enddata
\tablecomments{Exoplanet-hosting system properties.  In the property references column (rightmost column), {\bf bold}, \underline{underlined}, and \emph{italicized} numerals denote discovery, semimajor axis, and orbital period references, respectively. The final number in the column in normal font provides the companion object's mass reference.  All distances are from the \citet{gai21}.  {\bf References.} (1) \citet{for09}; (2) \citet{mar00}; (3) \citet{but06}; (4) \citet{fis08}; (5) \citet{bou18}; (6) \citet{cri18}; (7) \citet{but04}; (8) \citet{sou10}; (9) \citet{ge06}; (10) \citet{gui19}; (11) \citet{bou05}; (12) \citet{boi09}; (13) \citet{hen00}; (14) \citet{cha00}; (15) \citet{may95}; (16) \citet{mar15}.  }
\tablenotetext{a}{Also known as HD 285968 b.}
\end{deluxetable}

\begin{deluxetable}{lccccc}
\tabletypesize{\scriptsize}
\tablecolumns{6}
\tablewidth{0pt}
\tablecaption{Observations List}
\tablehead{
	\colhead{Name}&
	\colhead{UT Date}&
	\colhead{File Date}&
	\colhead{Scan Range\tablenotemark{a}}&
	\colhead{Number} &
	\colhead{Time on}\\
	\colhead{}&
	\colhead{(yyyy mm dd)}&
	\colhead{(yyyy mm dd)}&
	\colhead{}&
	\colhead{of Scans}&
	\colhead{Source (hr)}
}
\startdata
GJ 176 & 2010 01 10 & 2010 01 09 & 01800-03500 & 6 & 1.00\\
& 2010 12 19 & 2010 12 18 & 12500-13100{*} & 3 & 0.50\\
& 2010 12 19 & 2010 12 19 & 00000-00700 & 2 & 0.33\\
HD 46375 & 2010 01 10 & 2010 01 09 & 03600-04800{*} & 5 & 0.83\\
& 2010 01 10 & 2010 01 10 & 00000-00400 & 1 & 0.17\\
55 Cnc & 2011 01 02 & 2011 01 02 & 02700-05000 & 8 & 1.33\\
GJ 436 & 2010 01 06 & 2010 01 06 & 01800-03500 & 6 & 1.00\\
HD 102195 & 2010 01 07 & 2010 01 07 & 00000-01700 & 6 & 1.00\\
HD 189733\tablenotemark{b} & 2011 09 07 & 2011 09 06 & 00000-03500 & 12 & 2.00\\
HD 209458 & 2011 07 20 & 2011 07 20 & 04100-05800 & 6 & 1.00\\
& 2011 09 07 & 2011 09 06 & 03600-05500 & 7 & 1.17\\
51 Peg & 2011 07 19 & 2011 07 18 & 04100-06100 & 7 & 1.17\\
& 2011 09 07 & 2011 09 06 & 05700{*} & 1 & 0.17\\
& 2011 09 07 & 2011 09 07 & 00000-01600 & 5 & 0.83\\
\enddata
\tablecomments{Characteristics of AO data sets acquired and analyzed to search for star-planet interactions.}
\tablenotetext{a}{The scan range column denotes the continuous sequence of on-source, calibration-on, and calibration-off scans that focused on the listed target on a given day.  For example, file a2471.20101218.b0s1g0.12500.fits provides the on-source data to create the dynamic spectra from the first, lowest-frequency Mock spectrometer (b0) during the GJ 176 observation (scan 12500) that occurred on 2010 December 19 (Figure 1, left).  Asterisks in the scan range column denote an observing session that immediately continued into the following day.}
\tablenotetext{b}{Data from this observing session was presented in \citet{rou19} and \citet{rl19}.}
\end{deluxetable}

\begin{deluxetable}{llllllll}
\tabletypesize{\scriptsize}
\tablecolumns{6}
\tablewidth{0pt}
\tablecaption{Survey Detection Results}
\tablehead{
	\colhead{Object}&
	\colhead{Mass}&
	\colhead{Flux Density}&
	\colhead{$\nu$L$_{\nu}$}&
	\colhead{$\nu$L$_{\nu}$}&
	\colhead{$\nu$L$_{\nu}$\,\tablenotemark{a}}\\
	\colhead{}&
	\colhead{(M$_{J}$)}&
	\colhead{Limit (mJy)}&
	\colhead{(ergs s$^{-1}$)}&
	\colhead{(log L$_{\odot}$)}&
	\colhead{(log L$_{J,rad}$)}
}
\startdata
GJ 176 b & 0.0264 & $<$1.065 & $<$5.126$\times$10$^{23}$ & $<$-9.874 & $<$5.796\\
HD 46375 b & 0.226 & $<$1.134 & $<$5.282$\times$10$^{24}$ & $<$-8.862 & $<$6.809\\
55 Cnc e & 0.027 & $<$0.984 & $<$8.336$\times$10$^{23}$ & $<$-9.663 & $<$6.007\\
$\phantom{55 Cncx}$b & 0.8036 & $<$0.984 & $<$8.336$\times$10$^{23}$ & $<$-9.663 & $<$6.007\\
$\phantom{55 Cncx}$c & 0.1611 & $<$0.984 & $<$8.336$\times$10$^{23}$ & $<$-9.663 & $<$6.007\\
$\phantom{55 Cncx}$f & 0.1503 & $<$0.984 & $<$8.336$\times$10$^{23}$ & $<$-9.663 & $<$6.007\\
$\phantom{55 Cncx}$d & 3.12 & $<$0.984 & $<$8.336$\times$10$^{23}$ & $<$-9.663 & $<$6.007\\
GJ 436 b & 0.0737 & $<$1.047 & $<$5.352$\times$10$^{23}$ & $<$-9.856 & $<$5.815\\
HD 102195 b & 0.46 & $<$1.302 & $<$5.999$\times$10$^{24}$ & $<$-8.806 & $<$6.864\\
HD 189733 b\tablenotemark{b} & 1.13 & $<$1.158 & $<$2.421$\times$10$^{24}$ & $<$-9.200 & $<$6.470\\
HD 209458 b & 0.714 & $<$1.155 & $<$1.431$\times$10$^{25}$ & $<$-8.429 & $<$7.242\\
51 Peg b & 0.46 & $<$2.067 & $<$2.664$\times$10$^{24}$ & $<$-9.159 & $<$6.512\\
\hline
HR 8799\tablenotemark{c} e & 9.6 & $<$1.044 & $<$9.325$\times$10$^{24}$ & $<$-8.615 & $<$7.056\\
$\phantom{HR 8799x}$d & 7.2 & $<$1.044 & $<$9.325$\times$10$^{24}$ & $<$-8.615 & $<$7.056\\
$\phantom{HR 8799x}$c & 7.2 & $<$1.044 & $<$9.325$\times$10$^{24}$ & $<$-8.615 & $<$7.056\\
$\phantom{HR 8799x}$b & 5.8 & $<$1.044 & $<$9.325$\times$10$^{24}$ & $<$-8.615 & $<$7.056\\
\enddata
\tablecomments{Detection limits and radio luminosities of target systems.}
\tablenotetext{a}{Luminosity in terms of the average power output of Jupiter's ECM-induced decametric (DAM) during maximum solar activity (as opposed to average power or peak power), or $L_{J,rad}=8.2\times$10$^{17}$ erg~s$^{-1}$ \citep{zar04}.}
\tablenotetext{b}{Radio emission flux density detection limit presented in \citet{rou19}, with luminosity updated due to revised distance to HD 189733 \citep{gai21}.}
\tablenotetext{c}{Radio emission flux density detection limits for all HR 8799 exoplanets as reported in our first ultracool dwarf radio survey \citep{rou13}.}
\end{deluxetable}

\begin{deluxetable}{lllllll}
	\tabletypesize{\scriptsize}
	\tablecolumns{7}
	\tablewidth{0pt}
	\tablecaption{Orbital Phase Coverage of Radio Observations}
	\tablehead{
		\colhead{Object}&
		\colhead{Period}&
		\colhead{T$_{t}$}\tablenotemark{a}&
		\colhead{Obs Start}&
		\colhead{Obs End}&
		\colhead{Orbital Phases}&
		\colhead{Orbital Refs}\\
		\colhead{}&
		\colhead{(d)}&
		\colhead{(d)}&
		\colhead{(d)}&
		\colhead{(d)}&
		\colhead{Monitored}&
		\colhead{}
	}
	\startdata
	GJ 176 b & 8.7836 & 54399.790{*} & 55206.590 & 55206.634 & 0.853-0.858 & \underline{\bf 1}\\
	\phantom{GJ 176xb} & \phantom{8.7836} & \phantom{54399.79000} & 55549.655 & 55549.691 & 0.910-0.915& \\
	HD 46375 b & 3.023573 & 51071.357$\dagger$ & 55206.639 & 55206.683 & 0.681-0.695 & \underline{\bf 2}\\
	55 Cnc e & 0.7365474 & 57063.210{*} & 55563.752 & 55563.810 & 0.207-0.287 & \underline{\bf 3}\\
	GJ 436 b & 2.64389524 & 51549.560\phantom{*} & 55202.865 & 55202.909 & 0.789-0.806 & {\bf 4},\underline{2}\\
	HD 102195 b & 4.1139 & 53731.701$\dagger$ & 55203.855 & 55203.898 & 0.849-0.859 & {\bf 5},\underline{2}\\
	HD 189733 b\tablenotemark{b} & 2.218575 & 53629.389\phantom{*} & 55811.509 & 55811.598 & 0.568-0.608 & {\bf 6},\underline{7}\\
	HD 209458 b & 3.52474859 & 52854.830\phantom{*} & 55762.785 & 55762.829 & 0.011-0.023 & {\bf 4},\underline{2}\\
	\phantom{HD 209458 b} & \phantom{3.5247486} & \phantom{52854.830} & 55811.605 & 55811.660 & 0.861-0.877 & \\
	51 Peg b & 4.230785 & 50001.886$\dagger$ & 55761.784 & 55761.835 & 0.425-0.438 & \underline{\bf 2}\\
	\phantom{51 Peg b} & \phantom{4.230785} & \phantom{50001.886} & 55811.658 & 55811.709 & 0.214-0.226 & \\
	\enddata
	\tablecomments{Orbital phase coverage of the AO radio observations for systems with exoplanet targets that have orbital periods of $<$10 days.  Orbital periods are reported to maximum precision in the reference.  In the orbital references column, {\bf bold} and \underline{underlined} text denote orbital period and transit time (T$_{t}$) references, respectively. Transit times, observation start, and observation end dates (columns 3-5) are reported in HJD (or BJD) -2,4000,000.\\
	{\bf References.} (1) \citet{for09}; (2) \citet{but06}; (3) \citet{bou18}; (4) \citet{sou10}; (5) \citet{gui19}; (6) \citet{bou05}; (7) \citet{far17}.}
	
	\tablenotetext{a}{The T$_{t}$s are measured or estimated ($\dagger$), depending upon the reference. Asterisks ({*}) denote values that are in BJD.}
	\tablenotetext{b}{Orbital phase coverage from \citet{rou19}.}
\end{deluxetable}

\begin{figure}
	\centering
	\includegraphics[trim = 00mm 0mm 0mm 0mm, clip, width=0.8\textwidth,angle=0]{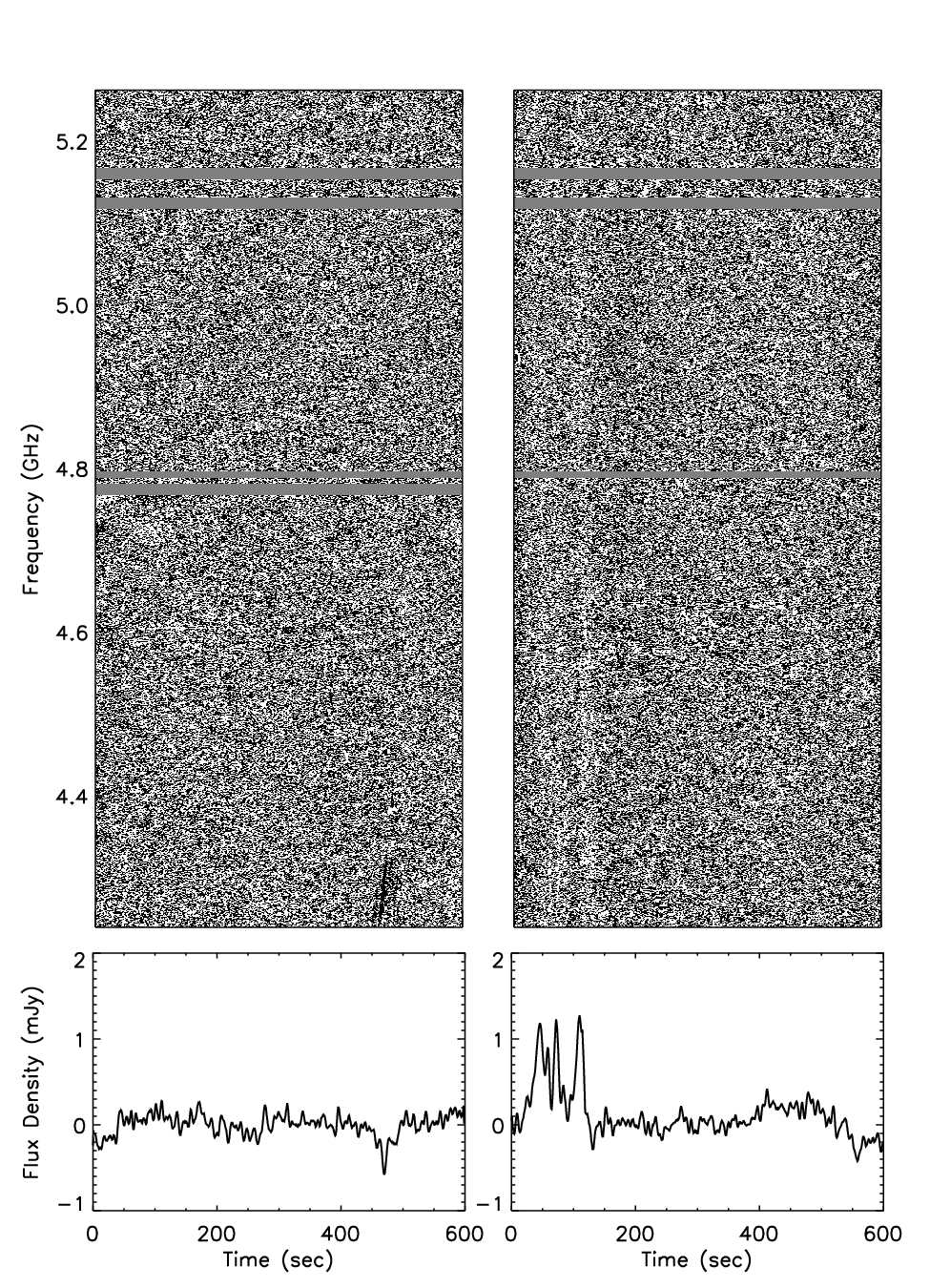}
	\caption{Stokes V dynamic spectra and bandpass-integrated time series from our exoplanet (left) and ultracool dwarf (right) radio surveys at AO.  Bandpass corrections, which consist of applying a polynomial fit from the central $\sim$80\% of each spectral window to the $\sim$10\% on each edge and renormalizing the flux density, mitigate subband discontinuities (scalloping) in sensitivity. Horizontal gray bars near 4.8 and 5.15 GHz represent the excision of strong RFI.  Even though a patch of left circularly polarized emission occurs at $t\sim$425 s in a dynamic spectrum from GJ 176 at left, follow-up analysis revealed that this feature is also strongly linearly polarized and appears in multiple data sets of different objects.  The emission is therefore an RFI artifact. At right is a real, $\sim$70\% circularly polarized radio flare from the T6.5 ultracool dwarf 2MASS J10475385+2124234 \citep{rou12,rou13}.}
\end{figure}

\begin{figure}
	\centering
	\includegraphics[width=0.9\textwidth,angle=0]{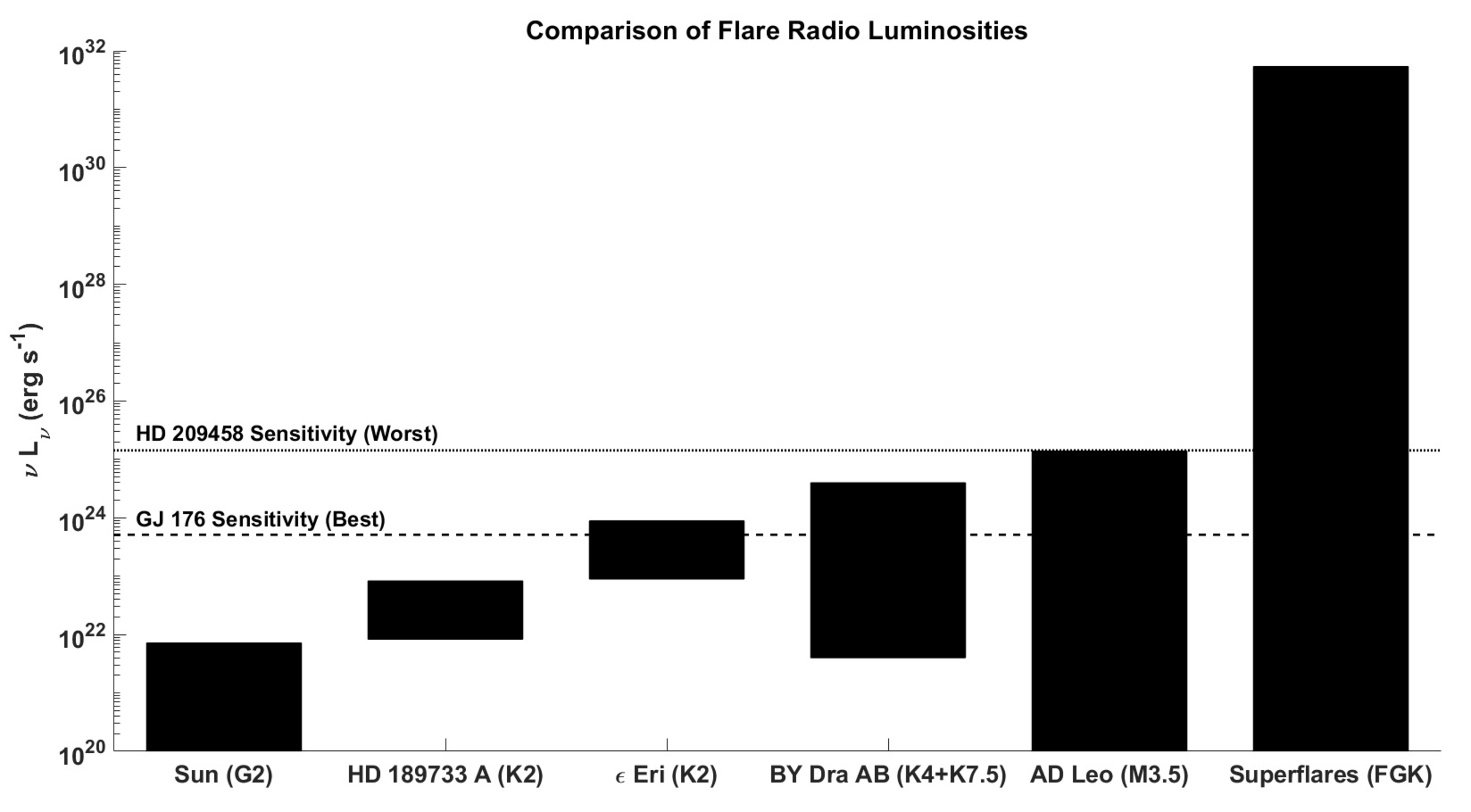}
	\caption{Survey $\nu L_{\nu}$ radio luminosity detection limits compared to known stellar flare luminosities.  For each object, the spectral type(s) are given in parentheses.  The flare radio luminosity ranges for HD 189733 A, $\epsilon$ Eri, and the interacting binary BY Dra AB are calculated from their X-ray flare luminosity ranges via the G\"{u}del-Benz relationship \citep{ben10,rou19}.  While these emissions are generated by incoherent gyrosynchrotron flaring, flares observed at AD Leo are created by the ECM process and are more readily detectable due to their coherence and beaming \citep{ste01}.  We establish a maximum luminosity for the ``superflares'' category using the 15 day $B$- and $V$-band flare observations of the F8 IV star 5 Serpentis, although it is nontrivial to relate the optical properties of the flares to those at radio wavelengths \citep{sch00,pri14}.  The dashed and dotted lines mark the detection thresholds of our most and least sensitive observations, those at GJ 176 and HD 209458, respectively, relative to these flare energies.}
\end{figure}

\begin{figure}
	\centering
	\includegraphics[width=0.8\textwidth,angle=0]{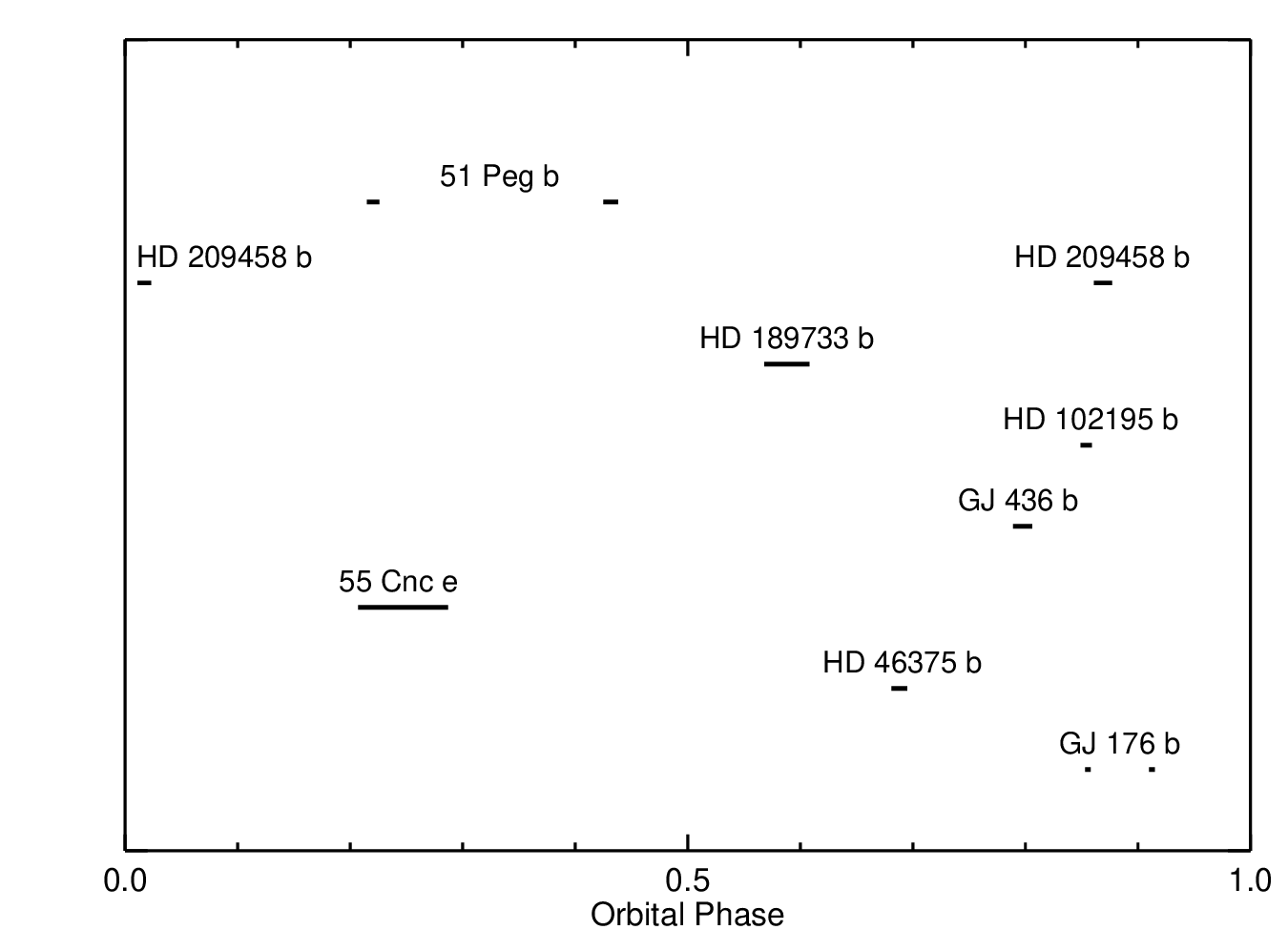}
	\caption{Orbital phase coverage of hot exoplanets ($P\leq$10 d) to search for magnetic star-planet interactions. Maximum orbital phase coverage occurs for 55 Cnc e, ($\Delta\phi_{orb}$=0.08).  GJ 176 b, HD 209458 b, and 51 Peg b each have two observing epochs depicted.  The orbital phase coverage of HD 189733 b occurs during phases where previous investigators suggested that they found increased stellar activity induced by potential magnetic star-planet interactions (ROME I) and is therefore significant.}	
\end{figure}

\end{document}